# Accurate near wall steady flow field prediction using Physics Informed Neural Network (PINN)

Vinothkumar Sekar, Qinghua Jiang, Chang Shu* and Boo Cheong Khoo
*Department of Mechanical Engineering,*
*National University of Singapore,*
*9 Engineering Drive 1, Singapore 117575.*

**Abstract**

In this paper, Physics Informed Neural Network (PINN) is explored in order to obtain flow predictions near the wall region accurately with measurements (or sampling points) away from the wall. Often, in fluid mechanics experiments, it is difficult to perform velocity measurements near the wall accurately. Therefore, the present study reveals a new and elegant approach to recover the flow solutions near the wall. Laminar boundary layer flow over a flat plate case is considered for this study in order to explore the ability of PINN to accurately predict the flow field. All the required sampling data for this study is obtained from CFD simulations. A wide range of Reynolds number cases from Re=500 to 100000 has been investigated. First, using PINN, the boundary layer solution is obtained with three different types of boundary conditions. Further, the influence of the location of the sampling points on the accuracy is analysed. From the velocity profiles and the skin friction coefficient distribution, it is clear that PINN results are reasonably accurate near the wall with only a few sampling points away from the wall. This approach has potential application in experiments to obtain the near wall solutions accurately with measurements away from the wall.

**Key Words:** Physics Informed Neural Network (PINN), Boundary Layer Flow, Computational Fluid Dynamics (CFD)

## 1. Introduction

Obtaining flow field is an essential task in fluid mechanics to find the flow properties, study the boundary layer and flow separation, and study the wake flow and vortex interactions. In mechanical and aerospace industries, Computational Fluid Dynamics (CFD) techniques have been tremendously utilized to obtain the flow field to perform the design and analysis of flow devices and components. However, in the final stage of the design, experimental analysis is necessary in order to validate the design inferences obtained from CFD. In experiments, it is often difficult to measure the velocity components near the wall region, especially the information within the boundary layer. This is due to the fact that the boundary layer is a thin region with steep velocity gradient. Even if measurements are done, they may be relatively less accurate or erroneous. The thin thickness of the boundary layer makes it difficult to distribute measuring devices. In the case of probe type meters, they cannot be placed near the wall, and for particle type measurements, the number of particles reaching the near-wall region might be insufficient. In the meantime, most high-accurate experimental techniques are expensive, and their usage is only necessary in limited areas near the wall boundary. The quantities away from the wall can be measured easily, even with low-resolution experimental techniques. Hence, it would be attractive to develop a method that can recover the near-wall flow field accurately, probably using the solution that is away from the wall. In this study, Machine Learning techniques are attempted to fulfill this requirement. In this study, CFD data is utilized as the sampling data (true data) to train the networks and validate the solutions obtained from the networks.

---

* Author to whom correspondence should be addressed. **Electronic mail:** mpeshuc@nus.edu.sg

In the literature, several attempts have been made to perform flow field prediction over various geometries using ML techniques. Guo et al. (2016) utilized CNN based ML techniques to predict the velocity field over several geometrical shapes and achieved an accuracy of about 98%. Jin et al. (2018) proposed $C_p - u$ model based on CNN to predict the unsteady velocity field around a circular cylinder from the measured pressure distributions. Sekar et al. (2019) utilized CNN combined with MLP to predict the flow field over various airfoils accurately. Bhatnagar et al. (2019) used CNN to predict the flow field over unseen airfoil geometries for turbulent flow. Thuerey et al. (2020) used a U-net architecture based on CNN to predict the flow field over a large number of airfoils. Further, several applications of machine learning in fluid mechanics are illustrated in Li et al., (2021), Liu et al., (2020) and Kashefi et al., (2020). The mentioned flow prediction models are purely based on input data and do not involve constraints from the governing equations.

Although being an attractive exploration, the purely data-driven ML models require a large number of sampling points distributed all over the domain to predict the flow field accurately. With fewer sampling points or in the case where the measurements are not possible in certain regions, such models fail to predict accurate solutions. An effective resolution of this issue is to introduce constraints of conservation laws, which yields the Physics Informed Neural Network (PINN) (Raissi et al., 2017a, 2017b & 2019a). In this method, a few sampling points are sufficient to obtain an accurate solution of the full domain. Another merit of PINN is its ability to efficiently handle ill-posed problems. In CFD, to obtain an accurate solution, the problem has to be well-posed with appropriate boundary conditions. It is rather complex and difficult to solve an ill-posed problem in CFD as it does not have a straightforward solution. On the contrary, the PINN can obtain an accurate solution even with ill-posed conditions. It is possible to obtain a solution even with certain flow quantities missing on the boundary, as easy as well-posed problems. This is one major advantage of PINN and makes it attractive for recovering solutions in certain regions where accurate measurements are difficult or even impossible.

Very limited explorations have been made to incorporate constraints of Navier-Stokes (N-S) equations into PINN. Raissi et al. (2017a, 2019) presented vortex shedding flow past an unsteady circular cylinder using PINN with 1% of the solution points as samples, which is probably the first N-S informed solution of PINN in the literature. Further, Raissi et al. (2018) recovered the velocity and the pressure fields of an unsteady circular cylinder with PINN by only utilizing spatio-temporal visualizations of a passive scalar (e.g., dye or smoke). In addition, Raissi et al. (2019) utilized PINN to obtain the full flow field and force coefficients of a cylinder undergoing vortex-induced vibration with only limited information on the velocity field. Jin et al. (2020) solved unsteady flow past a circular cylinder with PINN by utilizing a convergence-enhanced technique to obtain an accurate solution. These recent works show the successful application of PINN on fluid flow problems, but are mostly preliminary. In addition, it is also important to comprehensively validate the accuracy of the prediction, particularly in the near-wall region. Hence, the present work further explores the N-S constrained PINN through a detailed study on the laminar boundary layer flow over the flat plate.

The paper is outlined as follows. Section 2 gives the detail of vanilla Multilayer Perceptron Neural Network, Physics Informed Neural Network and the solution methodologies. Further, it describes the database preparation details, network configurations and its training details. Section 3 explains the results in detail for the boundary layer solution along a flat plate by PINN with different boundary conditions, internal sampling points and for a wide range of Reynolds numbers. Concluding remarks are given in Section 4.

## 2. Methodology

### 2.1 Physics Informed Neural Network (PINN)

A Physics Informed Neural network (Raissi et al., 2017b, 2019a) is a special type of network that respects the physical conservation laws. Depending on the application, PINN can be constructed using any type of Neural Networks. The PINN utilized in this study is constructed based on MLP (Rosenblatt, 1957) architecture, as shown in Fig. 1. The PINN embeds the steady incompressible N-S equations into the constraints, although PINN can be constructed with any physical laws. Similar to any MLP, the first layer is known as the input layer and receives the user-defined input. The last layer (output layer) predicts the desired output. The steady flow network receives the computational domain coordinates $x, y$ as input, and predicts the flow solutions $p, u, v$ as output. In the case of plain MLP, the loss function is purely based on the mean squared error (MSE) of sampled data points, while the PINN brings the physics into the network prediction by embedding the residuals of the N-S equation system in the form of loss function along with MSE, as shown in Fig. 1.

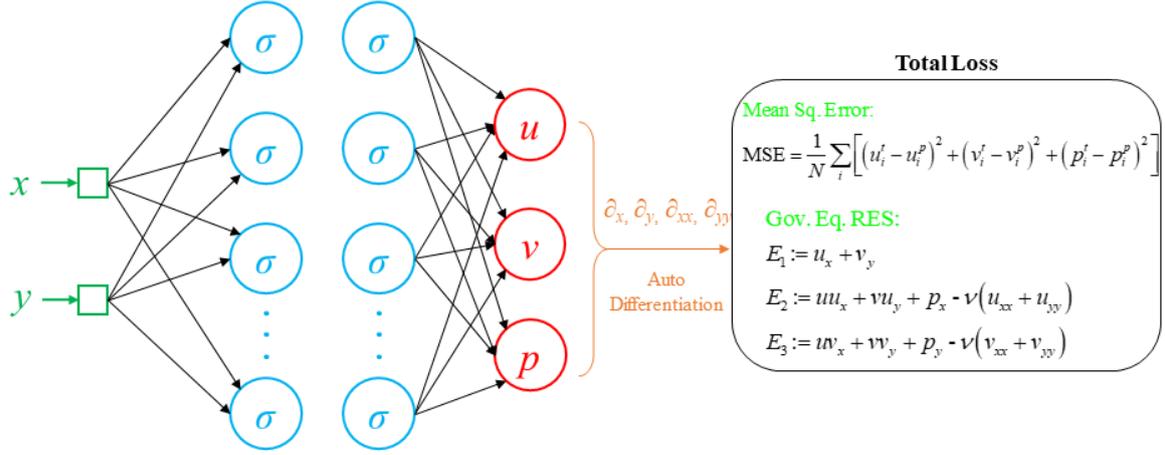

Figure 1 A typical MLP based PINN for flow field prediction

Let us consider the two dimensional (2D) steady incompressible N-S equations as follows.

$$\begin{aligned} u_x + v_y &= 0 \\ uu_x + vu_y &= -p_x + \nu(u_{xx} + u_{yy}) \\ uv_x + vv_y &= -p_y + \nu(v_{xx} + v_{yy}) \end{aligned} \quad (1)$$

where $u$ and $v$ represent velocity components in $x$ and $y$ directions, respectively; $p$ represents pressure normalized by density and $\nu$ indicates kinematic viscosity. The residuals of the steady incompressible N-S equations can then be expressed as

$$\begin{aligned} E_1 &= u_x + v_y \\ E_2 &= uu_x + vu_y + p_x - \nu(u_{xx} + u_{yy}) \\ E_3 &= uv_x + vv_y + p_y - \nu(v_{xx} + v_{yy}) \end{aligned} \quad (2)$$

The MSE of the N-S equations (residuals of the N-S in loss function) can be represented as

$$MSE_{NS} = \frac{1}{N_{NS}} \sum_{i=1}^{N_{NS}} (E_1^2 + E_2^2 + E_3^2) \quad (3)$$

where $N_{NS}$ indicates the number of solution points used to compute residuals of the N-S equations. The residuals of the N-S equations are computed for internal points as well as boundary condition points. The MSE of the sampled internal data points (sampling points) can be defined as,

$$MSE_S = \frac{1}{N_S} \sum_{i=1}^{N_S} (u_i^t - u_i^p)_S^2 + (v_i^t - v_i^p)_S^2 + (p_i^t - p_i^p)_S^2 \qquad (4)$$

and the MSE of the boundary condition points can be defined as,

$$MSE_B = \frac{1}{N_B} \sum_{i=1}^{N_B} (u_i^t - u_i^p)_B^2 + (v_i^t - v_i^p)_B^2 + (p_i^t - p_i^p)_B^2 \qquad (5)$$

where superscripts $t$ and $p$ represent true and predicted flow quantities, respectively. The boundary condition data points remain same for cases with or without internal sampling points. The MSE of the boundary condition points needs to be modified according to the type of boundary conditions utilized, as explored in Sec. 3.1. Therefore, the loss function for the PINN with internal sampling points can be defined as

$$Loss_{PINN} = MSE_S + MSE_B + MSE_{NS} \qquad (6)$$

The PINN network is trained such that the loss function defined in Eqn. 6 is minimized. One important step in PINN is the calculation of the required solution gradients with respect to the input variables. These gradients are utilized in the loss function, as shown in Eqn. 2. Automatic differentiation (Abadi et al., 2015) is utilized to obtain gradients efficiently. In addition, the automatic differentiation technique does not require mesh or cell details, and just specifying the points is sufficient to obtain gradients. In this way, PINN forms a special type of network and can be utilized in many applications where the conventional NNs cannot be employed.

**2.2 CFD Database**

In this study, the CFD solutions are considered as the true data in order to compare with the solution obtained from the PINN. The CFD simulation of the flat plate boundary layer is carried out using OpenFOAM solver (Weller et al., 1998, Jasak et al., 2007). Simulation data are obtained for several cases with Reynolds numbers ranging from 500 to 100000. The Reynolds number is calculated based on the length of the flat plate ($L$), which spans 5 units. The utilized computational domain for the CFD is shown in Fig. 2, consisting of $X: [-1,5] \times Y: [0,3]$. The computational domain for PINN starts from the leading edge of the flat plate, i.e., $X: [0,5] \times Y: [0,3]$. The flow field information ($p, u, v$) is sampled at $X = 0$ from the CFD simulations, which forms the inlet boundary sampling points for PINN. Similarly, the flow field information at the outlet and on the wall are also sampled from CFD solutions, which form the outlet and the wall boundary sampling points for PINN, respectively. A total of 800 solution points are sampled at the boundary, which is used as the boundary condition for PINN. In addition to the boundary points, internal sampling points are considered as shown in Fig. 9, either away from the boundary layer or inside the boundary layer. As the BL thickness reduces with increase of Reynolds number, the points are adjusted such that the sampling points follow the same pattern as shown in Fig. 9 (b). The internal sampling has a total of 54 points ($= 9 \times 6$) spanning the BL. Apart from the sampling points, a total of 20000 mesh points are considered, as shown in Fig. 4, in order to obtain the solution of the N-S equations, which are the same as the mesh points used in CFD simulation. These points are utilized to compute the residuals of the N-S equations in the loss function. Therefore, for each Re case, the database consists of 54 internal sampling points, 800 boundary sampling points and 20000 solution points.

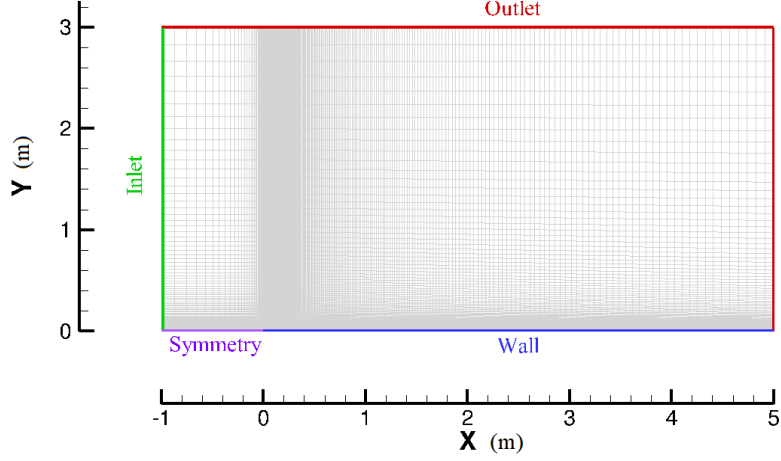

Figure 2 Mesh details utilized for obtaining CFD solution

**2.3 Training Details**

The network implementation and training are performed in Tensorflow (Abadi et al., 2015) open-source machine learning library. PINN training is an optimization process in which the weights and biases of the network are trained using a back-propagation algorithm (Rumelhart et al., 1988). The optimization is performed such that the defined total error of the PINN ($MSE_{PINN} = MSE_S + MSE_B + MSE_{NS}$) is minimized. A first-order gradient-based optimizer ADAM (Kingma et al., 2014), is used to train the network for the first 50000 iterations. The full batch training method is employed in which the weights are updated once after exposing full training data to the network. An initial learning rate (LR) of $1 \times 10^{-3}$ is specified. And if the error does not reduce for 1000 iterations during the training, LR is reduced by half. A minimum learning rate of $1 \times 10^{-7}$ is specified below which the LR is not reduced further. Once the training by ADAM is done, further training is performed using a second-order gradient-based optimizer known as L-BFGS (Byrd et al., 1995). The training ceases when the error no longer reduces. In most cases in our work, the ADAM is sufficient to get converged solutions, and the effect of L-BFGS is not significant.

# 3. Results and Discussion

The 2D incompressible laminar flat plate boundary layer flow is one of the famous benchmark case considered for validating the accuracy of the numerical solvers. The same is considered in this study with several scenarios for a Reynolds number range of $Re = 500$ to $100000$. Hyper-parameter selection is an important aspect in order to improve the performance of the network. Several architecture with varying hyper-parameters are considered for one of the case (BC-2 as mentioned in Sec. 3.1) and the convergence details of different architecture are shown in Table 1. For all the architectures, hyperbolic tangent activation function (Tanh) is considered for all the layers except for the last layer in which linear activation is considered. A base network of 8 layers with 50 neurons in each layer is considered initially. Further the neurons are increased to 100 and 120. The obtained MSE loss and the residuals of the N-S (N-S loss) for different networks are provided in Table 1. The convergence history of the different architecture are shown in Fig. 3. From the results, it is observed that increasing the architecture from $8 \times 50$ to $8 \times 100$ improves the accuracy slightly and further increasing to $8 \times 120$ does not show any improvement. From the results, taking the accuracy and computational time into account, the network with $8 \times 100$ architecture is chosen as a final architecture for further exploration.

| Hyper-parameter | MSE Loss | N-S Loss |
|---|---|---|
| $8 \times 50$ | $5.59 \times 10^{-5}$ | $1.14 \times 10^{-5}$ |
| $8 \times 100$ | $4.57 \times 10^{-5}$ | $6.52 \times 10^{-6}$ |
| $8 \times 120$ | $4.46 \times 10^{-5}$ | $7.21 \times 10^{-6}$ |

Table 1 Variation of hyper-parameter of PINN with error convergence

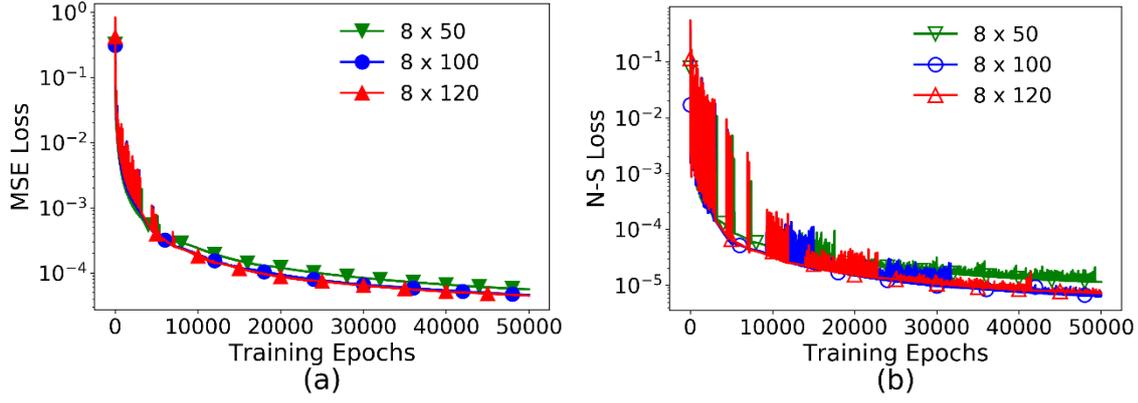

Figure 3 Convergence history for different hyper-parameter architecture of PINN (a) MSE loss (b) N-S loss

**3.1 Flow Solution without Internal Sampling points at Re=500**

At first, the ability of the PINN to solve the boundary layer flow with different types of boundary conditions (boundary sampling points) is explored. No internal sampling points are utilized, and only the boundary conditions are specified. This approach is similar to solving the fluid flow problems using any standard numerical technique by specifying boundary conditions. A total of 800 sampling points are used to specify the BCs, and a total of 20000 solution points are used for the PINN to obtain a solution and satisfy the N-S equations. The Reynolds number is explicitly specified in the momentum equation. Since there are no internal points used, the total loss for this case is the sum of boundary loss and the N-S residuals, as shown in Eqn. 7 as,

$$Loss_{PINN} = MSE_B + MSE_{NS} \qquad (7)$$

Here, three different types of BCs have been explored to check the solution accuracy of PINN. The three types of BCs are shown in Fig. 4, along with the computational domain and the solution points utilized. All the specified flow quantities ($u, v$ at the inlet for all BCs, $p$ on the wall, and $u, v$ at the outlet for BC-3) in the sampled boundary conditions are extracted from the CFD results. Boundary Condition Type-1 (BC-1) is a typical boundary condition applied in incompressible CFD cases, considered as well-posed conditions. In BC-1, velocity components $u$ and $v$ at the inlet are specified and zero-normal-pressure-gradient condition $(dp/dn = 0)$ is applied. On the wall surface, no-slip conditions ($u = 0, v = 0$) and zero-normal-pressure-gradient condition ($dp/dn = 0$) are implemented. At the outlet, zero-normal-velocity gradient conditions are applied for velocity, and pressure values are set to zero ($p = 0$). In BC-2, $u, v$ are specified at the inlet and on the wall, and $p$ is specified at the outlet. In BC-3, $p, u, v$ are specified at the inlet, on the wall, and at the outlet. For incompressible flows, BC-2 and BC-3 are considered as ill-posed boundary conditions (under-specified and over-specified), with which solutions cannot be obtained using conventional CFD techniques. According to the BCs utilized, the $MSE_B$ term in the loss function is modified. The $MSE_B$ term is modified as per Eqns. 8 to 10 for BC-1, BC-2, and BC-3, respectively.

$$MSE_{BC-1} = \left( \frac{1}{N} \sum_{i=1}^{N} (u_i^t - u_i^p)^2 + (v_i^t - v_i^p)^2 + (dp_{i_n})^2 \right)_{Inlet} + \left( \frac{1}{N} \sum_{i=1}^{N} (u_i^t - u_i^p)^2 + (v_i^t - v_i^p)^2 + (dp_{i_n})^2 \right)_{Wall} + \left( \frac{1}{N} \sum_{i=1}^{N} (du_{i_n})^2 + (dv_{i_n})^2 + (p_i^t - p_i^p)^2 \right)_{Outlet} \quad (8)$$

$$MSE_{BC-2} = \left( \frac{1}{N} \sum_{i=1}^{N} (u_i^t - u_i^p)^2 + (v_i^t - v_i^p)^2 \right)_{Inlet} + \left( \frac{1}{N} \sum_{i=1}^{N} (u_i^t - u_i^p)^2 + (v_i^t - v_i^p)^2 \right)_{Wall} + \left( \frac{1}{N} (p_i^t - p_i^p)^2 \right)_{Outlet} \quad (9)$$

$$MSE_{BC-3} = \left( \frac{1}{N} \sum_{i=1}^{N} (u_i^t - u_i^p)^2 + (v_i^t - v_i^p)^2 + (p_i^t - p_i^p)^2 \right)_{Inlet, Wall, Outlet} \quad (10)$$

The PINN training is performed, and the PINN solutions are obtained with each type of BCs. The contours of the obtained solutions using BC-1, BC-2, and BC-3, along with CFD results, are shown in Fig. 5. It is clear from the contours that all the obtained solutions show a good match with the CFD data. The comparison of the velocity profiles inside the boundary layer at $x = 1, 2, 3$, are shown in Fig. 6. The $u$-velocity profile obtained by the PINN agrees well with CFD for all types of BCs. However, the $v$-velocity profile shows a slight deviation from CFD for BC-1 and matches well with CFD for BC-2 and BC-3. The skin friction coefficient distribution over the wall is shown in Fig. 7, which is one of the important quantities to assess the near-wall accuracy of the solution. From Fig. 7, it is evident that although there exist slight oscillations, BC-2 and BC-3 show a reasonable match with the CFD, while BC-1 shows a greater deviation. The error convergence is shown in Fig. 8 and the averaged $L_2$ relative error (%) in the BL region is shown in Table 2. The PINN results in the case with BC-1 present a higher error, whereas the results with BC-2 and BC-3 show better accuracy (< 1% relative error). Compared to BC-2, BC-3 results in a slightly lower value of error in the BL regime. From the calculated residuals, it is clear that BC-2 has a better N-S residual convergence, as shown in Fig. 8 (b). In addition to the velocity components and the pressure, BC-1 involves its gradients to be satisfied at the boundaries, which probably makes it difficult for the optimizer to reduce the error further. BC-3 involves all the flow quantities $(p, u, v)$ to be specified at the boundaries, which makes the model to overfit the boundary sampling points than the N-S residual. But BC-2 involves the optimum flow quantities to be specified at the boundaries, which makes a balance between the boundary sampling points and the N-S residual. Hence, BC-2 shows better N-S residual convergence. Devising additional convergence-enhancement strategies or utilizing more efficient optimizers may improve the convergence further.

The obtained results give clear evidence that PINN is a powerful yet unconventional numerical technique in order to solve N-S equations. From the contours and velocity profile, the obtained results are reasonably accurate. The influence of BCs on the accuracy is not clearly visible in the solution contours but evident in the skin friction distribution. BC-2 is considered for further exploration because of its better convergence of the N-S residual.

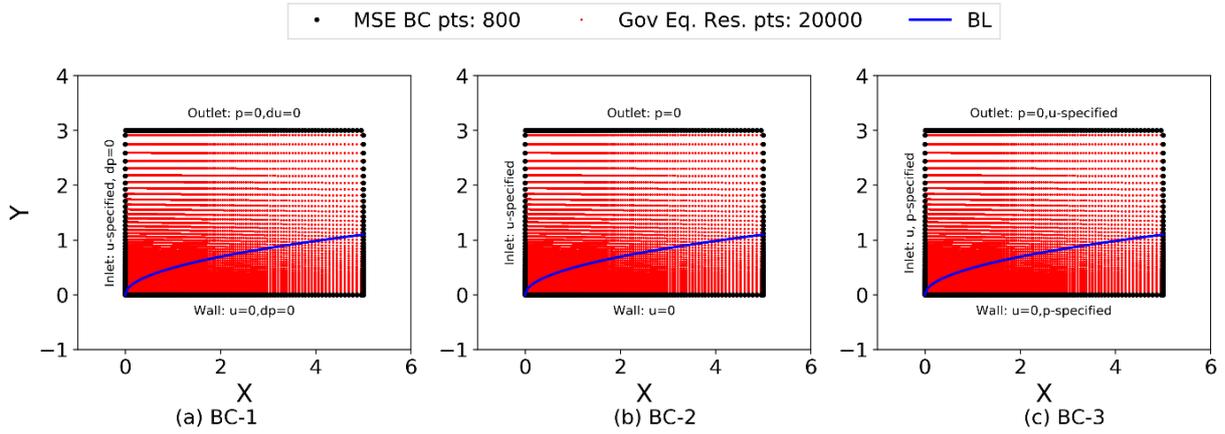

Figure 4  Boundary Condition Type -1,2,3 utilized to obtain PINN solution: (a) BC-1, (b) BC-2, (c) BC-3

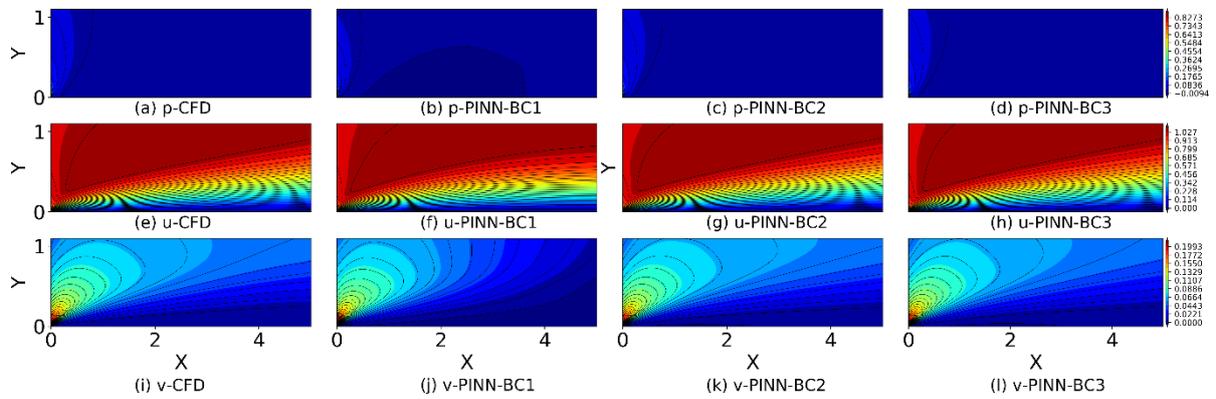

Figure 5 Comparison of contours of $p, u$ and $v$ for PINN (BC-1, BC-2 and BC-3) along with CFD: (a) p-CFD, (b) p-PINN-BC1, (c) p-PINN-BC2, (d) p-PINN-BC3 (e) u-CFD (f) u-PINN-BC1, (g) u-PINN-BC2, (h) u-PINN-BC3, (i) v-CFD, (j) v-PINN-BC1, (k) v-PINN-BC2, (l) v-PINN-BC3

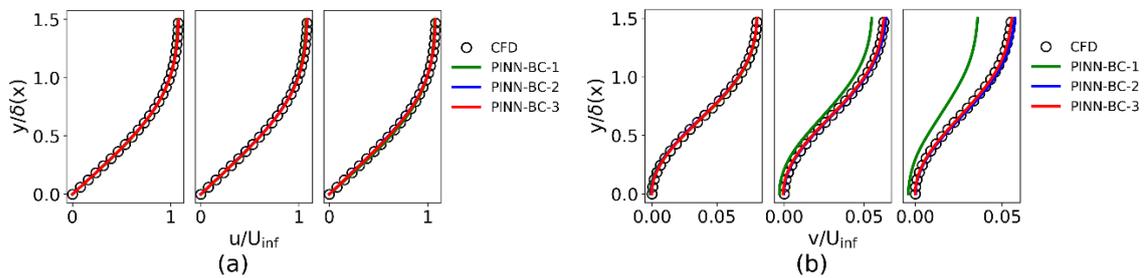

Figure 6 Comparison of velocity profiles for PINN (BC-1, BC-2, BC-3) along with CFD at $x = 1, 2, 3:$  (a) normalized u-velocity profile, (b) normalized v-velocity profile

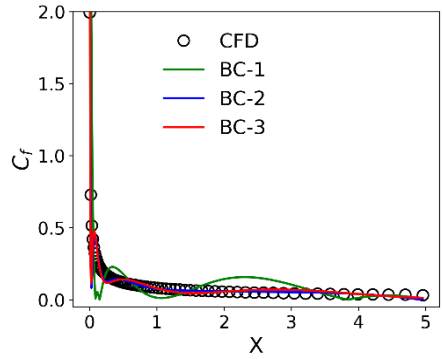

Figure 7 Comparison of skin friction coefficient distribution for PINN (BC-1, BC-2, BC-3) along with CFD

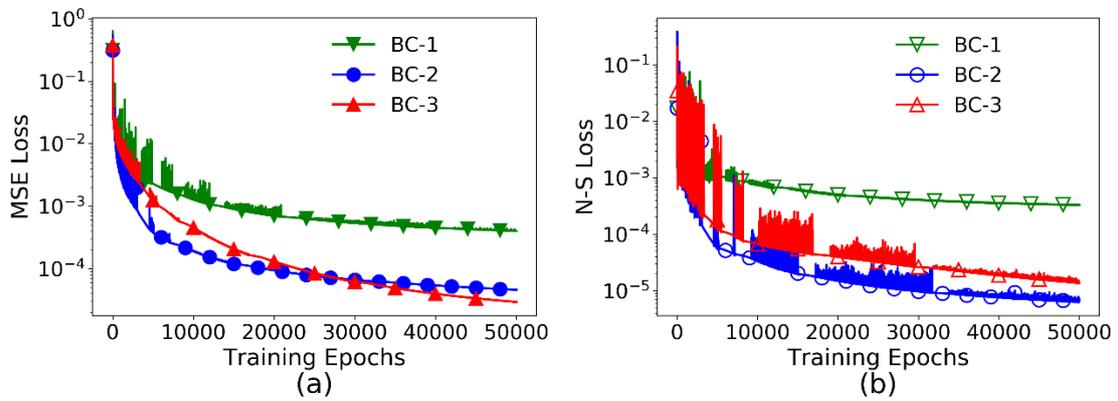

Figure 8 Comparison of losses for PINN (BC-1, BC-2, BC-3) (a) MSE Loss (b) N-S Loss

| Re=500 without internal sampling points | Average L2 relative error (%) in BL regime |
|---|---|
| BC-1 | 6.20 |
| BC-2 | 0.81 |
| BC-3 | 0.76 |

Table 2 Comparison of averaged L2 relative error (%) of PINN solution (BC-1, BC-2, BC-3) along with CFD

### 3.2 Flow Solution with Internal Sampling points at Re=500

In this case, the internal sampling points are utilized along with BC-2 in order to enhance the accuracy of the solution near the wall. This scenario is more realistic in experimental fluid mechanics, where the obtained measurements away from the wall can be directly used as sampling points. The effect of the location of the sampling points on the solution accuracy is analyzed. Two sets of sampling points are chosen: one set of sampling points (S-1) away from the boundary layer and the other set of sampling points within the boundary layer (S-2) but away from the wall. The sampling set consists of 54 (=9 × 6) points, as shown in Fig. 9. Since internal points are used, the total loss for this case is the sum of losses due to internal sampling, boundary sampling, and the N-S residuals, as shown in Eqn. 11 as,

$$Loss_{PINN} = MSE_S + MSE_B + MSE_{NS} \qquad (11)$$

The results obtained using PINN with a different set of sampling points are compared. The contours of the flow solution for two sampling cases, along with CFD comparison, are shown in Fig. 10. The velocity profiles within the boundary layer extracted at $x = 1, 2, 3$ are shown in Fig. 11, along with CFD results. The effect of sampling points is not visible in the contours or the velocity profiles in comparison with the previous case without sampling points. The skin friction coefficient distribution is shown in Fig. 12, along with CFD data. It is evident that the distribution of skin friction coefficient matches well with the CFD results in the case of S-2 (within the BL). For the S-1 (away from BL), there is no improvement in the results near the wall. This is intuitively true that distributing a few sampling points within the region of interest could improve the accuracy. The comparison of the convergence is shown in Fig. 13. The averaged $L_2$ relative error (%) is shown in Table 3 for both S-1 and S-2 cases, and the relative error in S-2 is slightly lower than that in S-1. Hence, BC-2, along with S-2 (sampling within BL), are utilized in the following analyses for further Reynolds number cases.

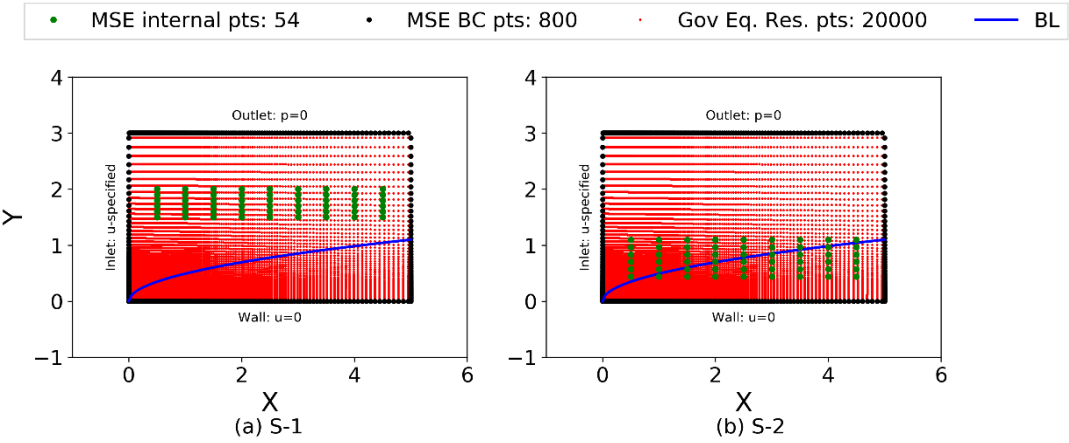

Figure 9 Distribution of internal sampling points for PINN (a) S-1 (b) S-2

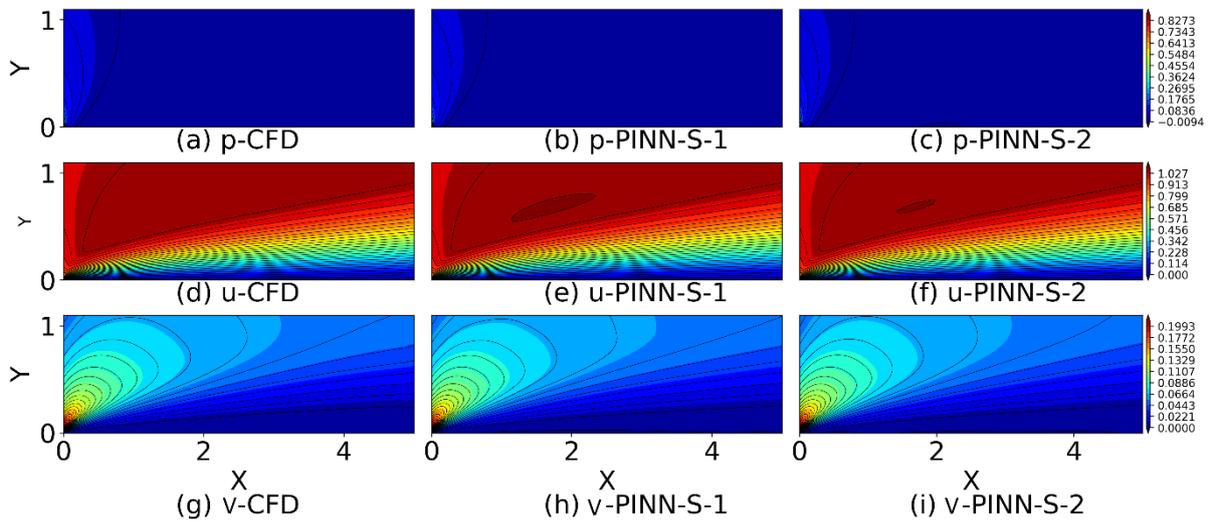

Figure 10 Comparison of contours of $p, u$ and $v$ for PINN (S-1, S-2) along with CFD: (a) p-CFD, (b) p-PINN-S-1, (c) p-PINN-S-2, (d) u-CFD, (e) u-PINN-S-1, (f) u-PINN-S-2, (g) v-CFD, (h) v-PINN-S-1, (i) v-PINN-S-2

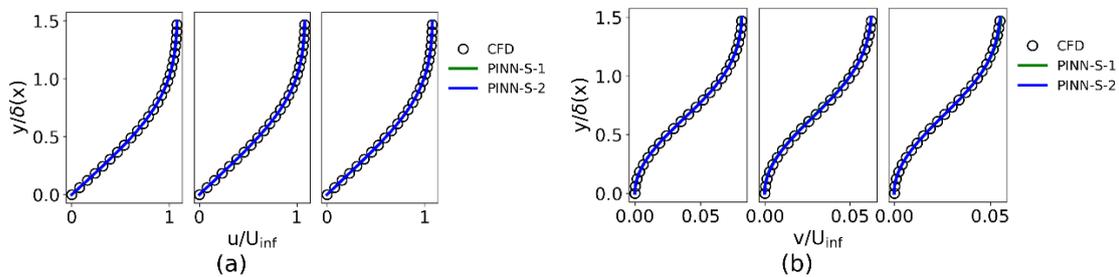

Figure 11 Comparison of velocity profiles for PINN (S-1, S-2) along with CFD at $x = 1, 2, 3$: (a) normalized u-velocity profile, (b) normalized v-velocity profile

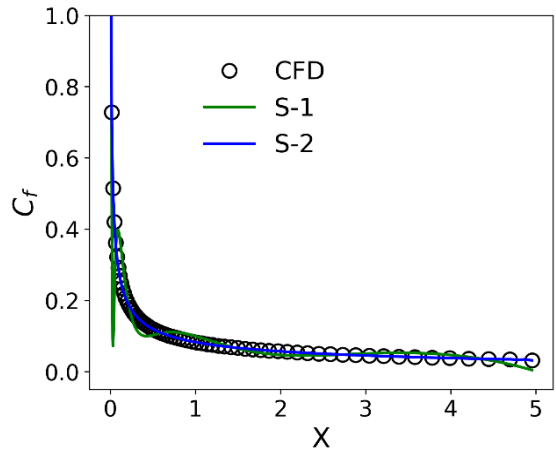

Figure 12 Comparison of skin friction coefficient distribution for PINN (S-1, S-2) along with CFD

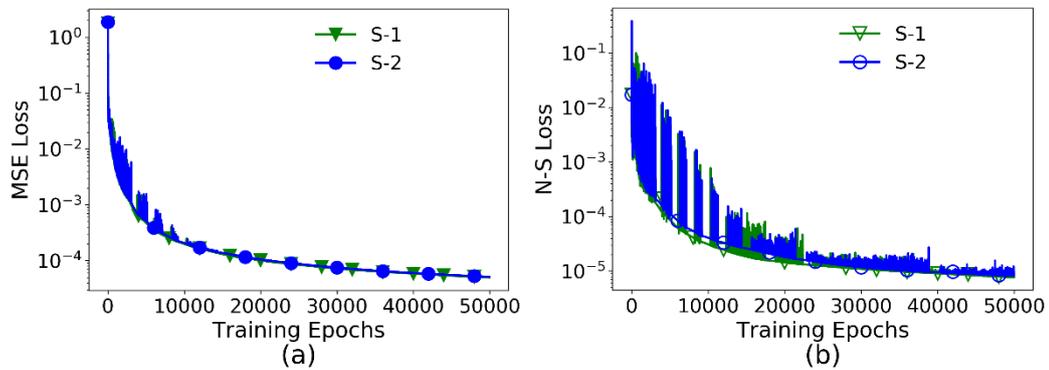

Figure 13 Comparison of losses for PINN (S-1, S-2) (a) MSE Loss (b) N-S Loss

| Re=500 with sampling points | Average L2 relative error (%) in BL regime |
|---|---|
| S-1 | 0.70 |
| S-2 | 0.65 |

Table 3 Comparison of averaged $L_2$ relative error (%) of PINN solution (S-1, S-2) along with CFD

### 3.3 Flow Solution with Internal Sampling points at Re=5000-100000

In this part, further tests are carried out to validate the PINN at Reynolds numbers from $Re = 5000$ to $100000$. The sampling within the BL (S-2), along with BC-2, is utilized for all the cases. As the BL thickness reduces with the increase of Re, the sampling points are adjusted so that the points are distributed in a similar pattern, as shown in Fig. 9 (b). For Re=5000, the velocity profiles, and skin friction distributions are given in Fig. 14, and Fig. 15, respectively, along with CFD results. From the results, it is evident that the obtained PINN solutions are reasonably accurate at this Reynolds number. Increasing the Reynolds number to Re=50000, the results are shown in Fig. 16, and Fig. 17. The PINN solutions are still in good agreement with the CFD results at this Reynolds number. Further, the results obtained at Re=100000 are presented in Fig. 18, and Fig. 19, respectively, for velocity profiles, and skin friction distributions. Deviations in the skin-friction coefficient distribution start to emerge at this Reynolds number. At $Re > 100000$, deviations of velocity profiles and skin friction distributions from the CFD results continue to increase, which is demonstrated by the quantitative comparisons of averaged $L_2$ relative error (%) in Table 4.

It is important to note that for all the cases within the range of Re=5000-100000, the same number of sampling points (54 points) is utilized. As the BL gets thinner, the sampling points are adjusted, such that more than half of the points are located inside the BL, as shown in Fig. 9 (b). Overall, with very few sampling points, PINN is able to recover the solutions near the wall region with reasonable accuracy, as can be seen from its recoveries of the skin friction distributions. The solution accuracy of the PINN reduces with the increase of the Reynolds number. And significant deviations from the CFD results are visible at Re=100000. Cases at Reynolds numbers higher than Re=100000 are not considered in this study as it is challenging to obtain good convergence and may need to consider turbulence effects. Often in fluid mechanics applications, the near-wall region is of primary interest, and it is difficult to perform accurate flow measurements near the wall. The PINN proposed in this study possesses the potential of recovering the near-wall flow solutions accurately with sparse measurements away from the wall region.

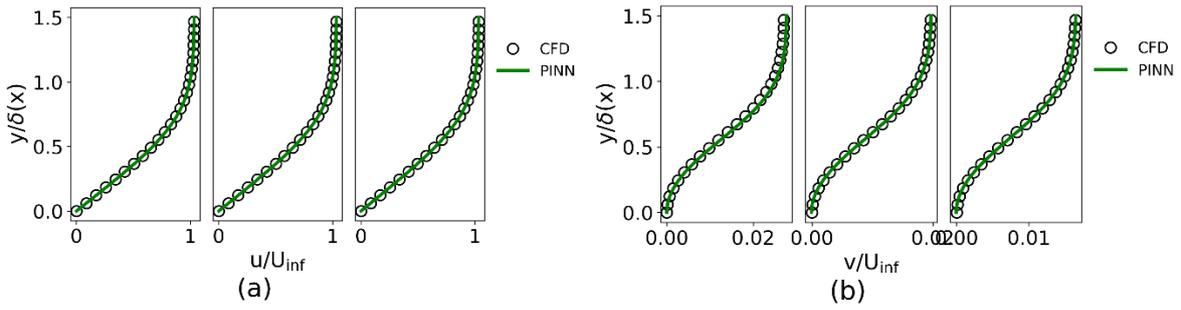

Figure 14 Comparison of velocity profiles for PINN for Re=5000 along with CFD at $x = 1, 2, 3:$ (a) normalized u-velocity profile (b) normalized v-velocity profile

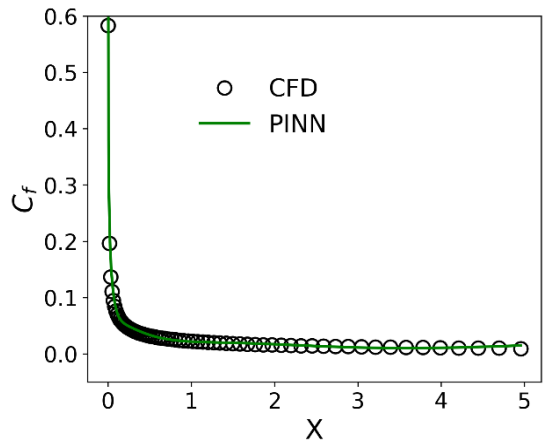

Figure 15 Comparison of skin friction coefficient distribution for PINN for Re=5000 along with CFD

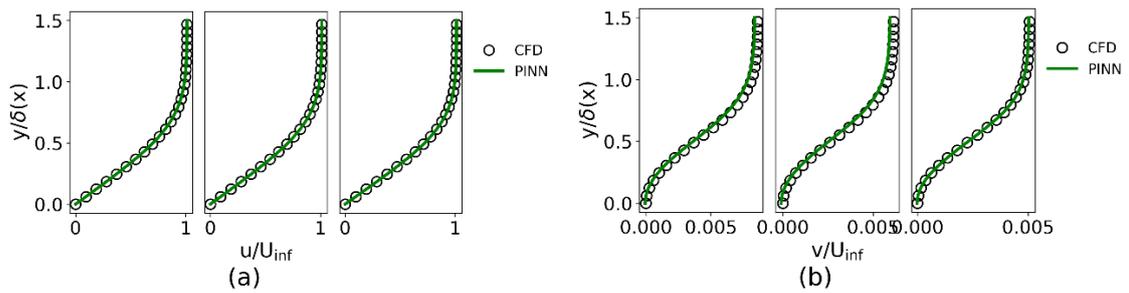

Figure 16 Comparison of velocity profiles for PINN for Re=50000 along with CFD at $x = 1, 2, 3:$ (a) normalized u-velocity profile, (b) normalized v-velocity profile

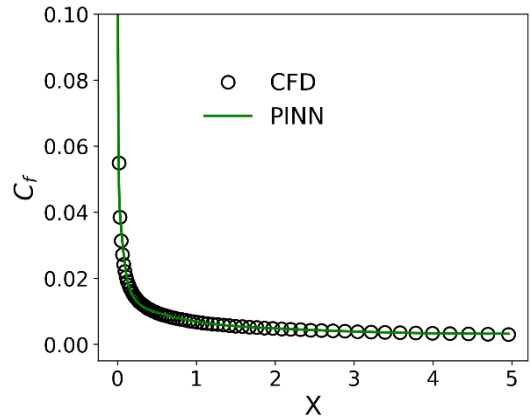

Figure 17 Comparison of skin friction coefficient distribution for PINN for Re=50000 along with CFD

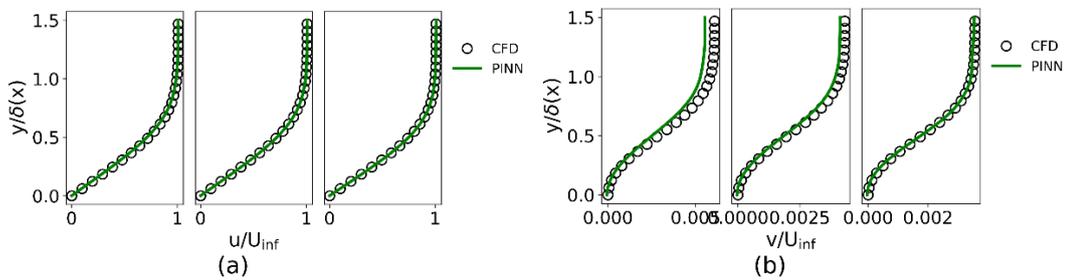

Figure 18 Comparison of velocity profiles for PINN for Re=100000 along with CFD at $x = 1, 2, 3:$ (a) normalized u-velocity profile, (b) normalized v-velocity profile

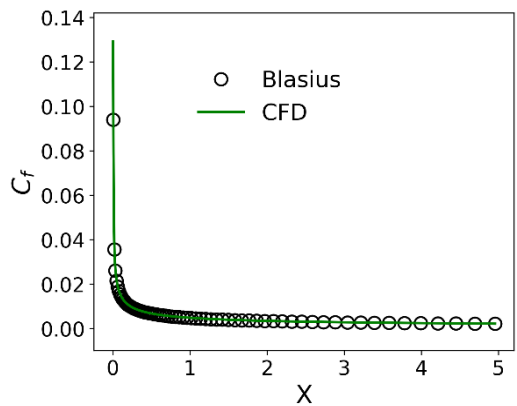

Figure 19 Comparison of skin friction coefficient distribution for PINN for Re=100000 along with CFD

| **Further Re cases** | **Averaged $L_2$ relative error (%) in BL regime** |
|---|---|
| Re=5000 | 6.79 |
| Re=50000 | 8.91 |
| Re=100000 | 20.52 |

Table 4 Comparison of averaged L$_2$ relative error (%) of PINN solution for Re=5000, 50000, and 100000 along with CFD

## 4. Conclusions

In this study, Physics Informed Neural Network (PINN) is explored in order to obtain accurate flow solutions near the wall. Often, in experiments, it is difficult to measure the velocity components near the wall accurately and velocity measurements away from the wall is relatively easier to obtain. Therefore, PINN is employed in order to analyse its ability to obtain solutions near the wall accurately with information away from the wall. Laminar flow over a flat plate boundary layer case is considered for this study. Solutions are obtained for the flat plate boundary layer case using PINN with several scenarios.

First, an attempt is made to obtain solution without any internal sampling points with three different boundary conditions (BC-1, BC-2 and BC-3) at Re=500. From the results, it is clear that PINN solution is reasonably accurate for Re=500 compared to CFD for BC-2 and BC-3. However, the skin-friction distribution shows up some oscillations in the distribution. Further, in order to improve the accuracy, internal sampling points are considered. Solutions are obtained using two sets of internal sampling points, one set away from the BL (S-1) and another set within the BL (S-2). It is observed that as the sampling points are introduced within the BL, the skin friction distribution matches accurately with the CFD data. This is intuitive that keeping fewer sampling points (measurements) away from the wall may help to recover the solution accurately near the wall. Also, it is noted that keeping sampling points too far away from the wall (away from BL) does not have any influence in the near wall solution.

Further, the solutions are obtained for cases from Re=5000 to 100000. As the Re increases, the solution accuracy of the PINN starts to deviate near the wall which is evident from the skin fiction distribution. Below Re=100000, the results seem to have a reasonable accuracy, and for Re=100000, the results show a significantly visible deviation from the CFD data. Hence, it is observed that for $Re > 100000$, it is challenging for PINN to obtain solutions near the wall accurately. The solutions may be improved further either by devising additional strategies or by introducing turbulence effects into account. In future, it would be interesting to improve the PINN's prediction for high Re cases, including turbulent cases.

## Acknowledgments

The authors are thankful to the National Supercomputing Centre (NSCC), Singapore for providing access to supercomputing facility. In addition, the first and second authors are thankful to National University of Singapore (NUS) and Ministry of Education (MOE), Singapore for research scholarship. Besides, the authors would like to thank Dr. Murali Damodaran of Temasek Laboratories, NUS for his valuable suggestions on the manuscript.

## Appendix

**A. Details on the CFD simulation**

In this study, the required CFD simulation of the flat plate boundary layer is carried out using OpenFOAM package (Weller et al., 1998, Jasak et al., 2007), using SIMPLE solver (Patankar, 1980). The Reynolds number is calculated based on the length of the flat plate ($L$), which spans 5 units. The utilized computational domain for the CFD is shown in Fig. 2, consisting of $X: [-1,5] \times Y: [0,3]$. The computational domain for PINN starts from the leading edge of the flat plate, i.e., $X: [0,5] \times Y: [0,3]$. CFD simulations are carried out for several Reynolds number from $Re = 500$ to $100000$. The obtained CFD results for $Re = 50000$ and $Re = 100000$ are validated with analytical Blasius solution (White, 2006). Fig. 20 shows the comparison velocity profile obtained from CFD with Blasius solution for Re=50000 and 100000. The comparison of skin-friction distribution obtained from CFD with Blasius solution is shown in Fig. 21. From the obtained velocity profile and skin-friction distribution, it is evident that the obtained CFD solution for the laminar boundary layer flow over a flat plate case is accurate.

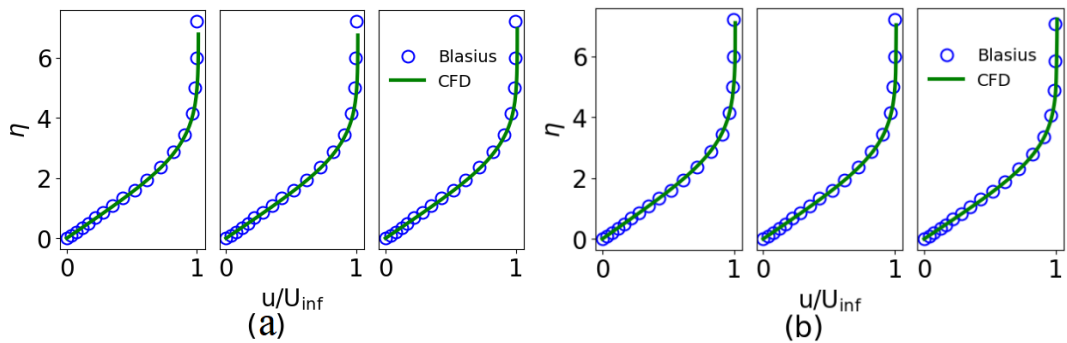

Figure 20 Comparison of CFD Velocity profile (at $x = 1, 2,$ and $3$) with Blasius solution for (a) Re=50000 and (b) Re=100000 (where $\eta = y\sqrt{U/\upsilon x}$)

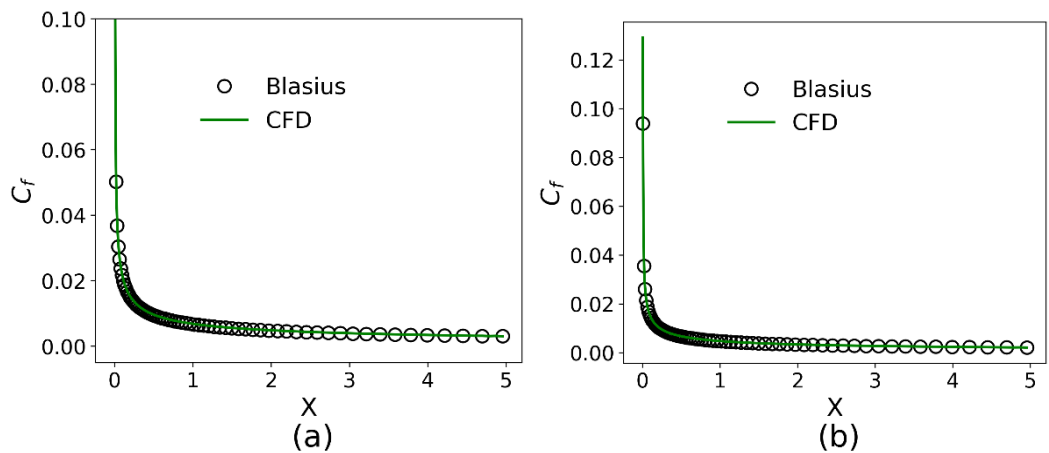

Figure 21 Comparison of CFD skin-friction distribution with Blasius solution for (a) Re=50000 and (b) Re=100000

### B. Comparison of PINN's contours along with CFD

The comparison of $p, u$ and $v$ contours obtained from PINN along with CFD for $Re = 5000, 50000$ and $100000$ are shown in Fig. 22, 23 and 24, respectively.

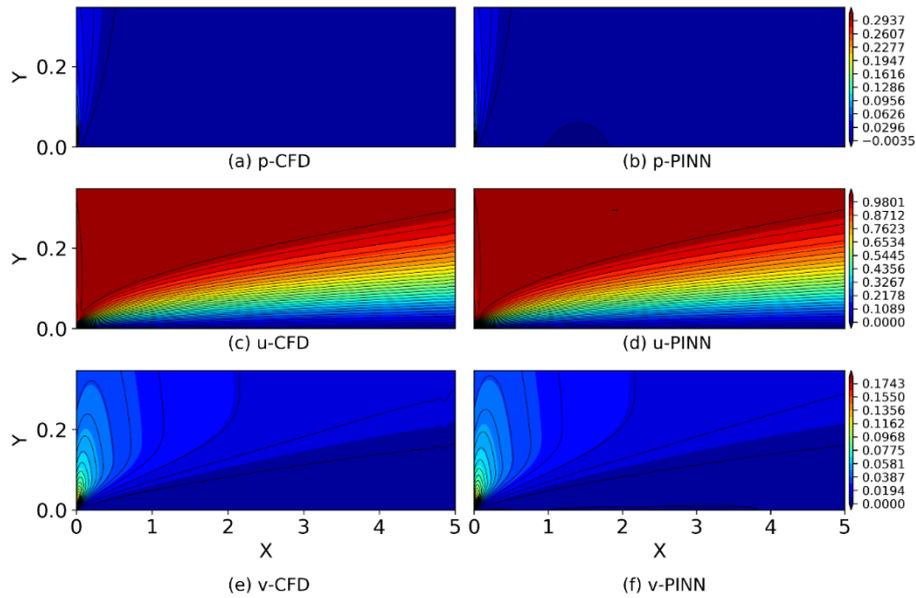

Figure 22 Comparison of contours of $p, u$ and $v$ for PINN for Re=5000 along with CFD: (a) p-CFD, (b) p-PINN, (c) u-CFD, (d) u-PINN, (e) v-CFD, (f) v-PINN

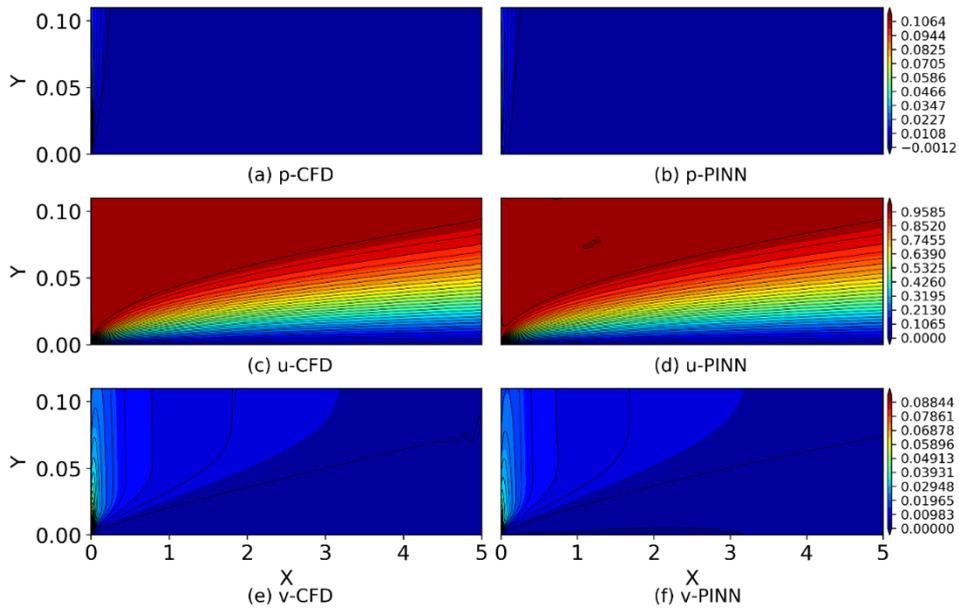

Figure 23 Comparison of contours of $p, u$ and $v$ for PINN for Re=50000 along with CFD: (a) p-CFD, (b) p-PINN, (c) u-CFD, (d) u-PINN, (e) v-CFD, (f) v-PINN

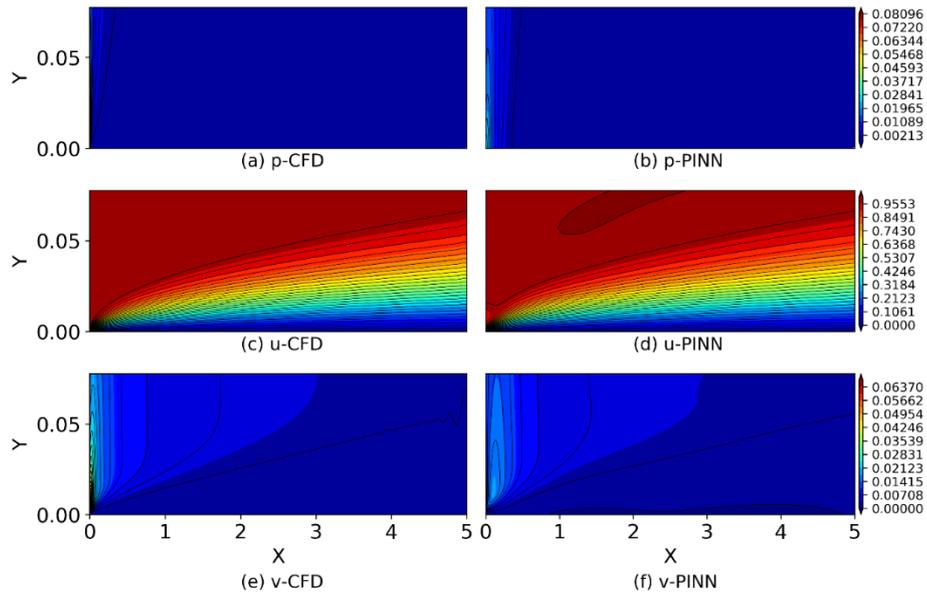

Figure 24 Comparison of contours of $p$, $u$ and $v$ for PINN for Re=100000 along with CFD: (a) p-CFD, (b) p-PINN, (c) u-CFD, (d) u-PINN, (e) v-CFD, (f) v-PINN

**C. Comparison of solution obtained from PINN with NN**

In order to understand the effectiveness of the PINN, the prediction results obtained with PINN and vanilla NN are compared for the case type of BC-2 without internal sampling points as mentioned in Sec. 3.1. The comparison of the pressure and velocity component contours are shown in Fig. 25. The extracted velocity profile comparisons are shown in Fig. 26. From the comparison, it is evident that PINN can obtain accurate solution from the given fewer sampling points whereas vanilla NN fails to predict a reasonable solution.

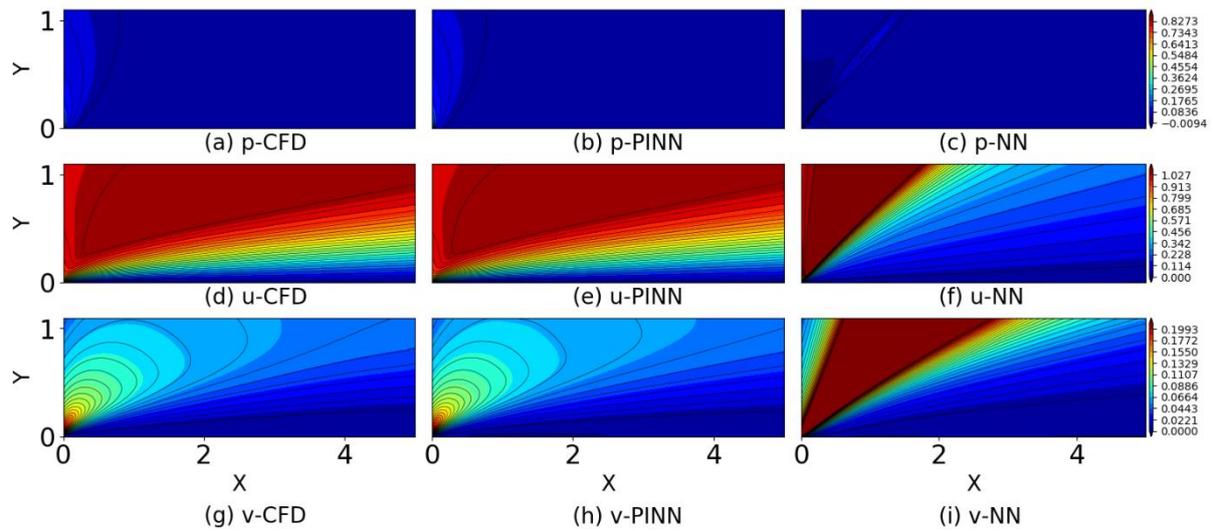

Figure 25 Comparison of contours for PINN and vanilla NN along with CFD: (a) p-CFD, (b) p-PINN (c) p-NN, (d) u-CFD, (e) u-PINN, (f) u-NN, (g) v-CFD, (h) v-PINN, (i) v-NN

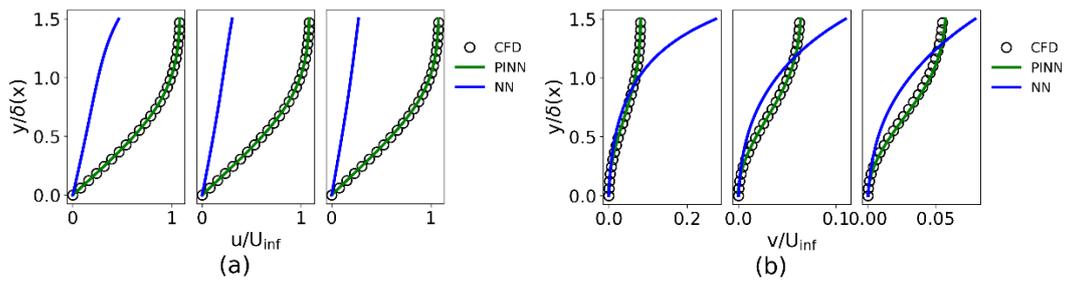

Figure 26 Comparison of velocity profiles for PINN and vanilla NN along with CFD: (a) normalized u-velocity profile, (b) normalized v-velocity profile

# References

Abadi, M. et al., TensorFlow: "Large-scale machine learning on heterogeneous systems," (2015). software available from tensorflow.org.

Bhatnagar, S., Afshar, Y., Pan, S. *et al.* Prediction of aerodynamic flow fields using convolutional neural networks. *Comput. Mech* **64,** 525–545 (2019). https://doi.org/10.1007/s00466-019-01740-0

Byrd, R.H., Lu, P., and J. Nocedal. "A Limited Memory Algorithm for Bound Constrained Optimization", (1995), SIAM Journal on Scientific and Statistical Computing, 16, 5, pp. 1190-1208.

Duraisamy, K., Iaccarino, G., and Xiao, H., "Turbulence modeling in the age of data," *Annual Review of Fluid Mechanics*, *51*, 357-377, (2019).

Guo, X., Li, W., and Iorio, F., "Convolutional Neural Networks for Steady Flow Approximation," In *Proceedings of the 22nd ACM SIGKDD International Conference on Knowledge Discovery and Data Mining* (KDD '16). ACM, New York, NY, USA, 481-490, (2016).

Hornik, K., Stinchcombe, M., and White, H., "Multilayer feedforward networks are universal approximators", Neural Networks, Vol. 2, Issue 5, (1989), Pages 359-366, ISSN 0893-6080, https://doi.org/10.1016/0893-6080(89)90020-8.

Jasak, H., Jemcov, A., and Tukovic, Z., "OpenFOAM: A C++ library for complex physics simulations," In International workshop on coupled methods in numerical dynamics ,Vol. 1000, pp. 1-20, IUC Dubrovnik Croatia, (2007).

Jin, X., and Cheng, P., and Chen, W-L., and Li, H., "Prediction model of velocity field around circular cylinder over various Reynolds numbers by fusion convolutional neural networks based on pressure on the cylinder," Physics of Fluids 30, 047105 (2018).

Jin, X., Cai, S., Li, H., & Karniadakis, G. E. (2020), NSFnets (Navier-Stokes Flow nets): Physics-informed neural networks for the incompressible Navier-Stokes equations, arXiv:2003.06496v1

Kashefi, A., Rempe, D., and Guibas, L.J., "A point-cloud deep learning framework for prediction of fluid flow fields on irregular geometries", Physics of Fluids 33, 027104 (2021). https://doi.org/10.1063/5.0033376

Kingma, D. P., and Ba, J., "Adam: A method for stochastic optimization," CoRR, Vol. abs/1412.6, (2014).

Li, Y., Chang, J., Wang, Z., and Kong, C., "An efficient deep learning framework to reconstruct the flow field sequences of the supersonic cascade channel", Physics of Fluids 33, 056106 (2021).

Ling, J., and Templeton, J., "Evaluation of machine learning algorithms for prediction of regions of high Reynolds averaged Navier Stokes uncertainty," Physics of Fluids, Vol. 27, No. 8, pp. 085103, (2015). https://aip.scitation.org/doi/abs/10.1063/1.4927765

Liu, B., Tang, J., Huang, H., and Lu, X-Y., "Deep learning methods for super-resolution reconstruction of turbulent flows", Physics of Fluids 32, 025105 (2020). https://doi.org/10.1063/1.5140772

Ma, M., Lu, J., and Tryggvason, G., "Using statistical learning to close two-fluid multiphase flow equations for a simple bubbly system," Physics of Fluids, Vol. 27, No. 9, pp. 092101, (2015). https://aip.scitation.org/doi/abs/10.1063/1.4930004

Maulik, R., San, O., Rasheed, A., and Vedula, P., "Data-driven deconvolution for large eddy simulations of Kraichnan turbulence," Physics of Fluids, Vol. 30, No. 12, pp. 125109, (2018). https://aip.scitation.org/doi/abs/10.1063/1.5079582

Patankar, S. V., "Numerical Heat Transfer and Fluid Flow", Taylor & Francis, (1980), ISBN 978-0-89116-522-4.

Raissi, M., Perdikaris, P. & Karniadakis, G. E. (2017a), Physics informed deep learning (part II): data-driven discovery of nonlinear partial differential equations. arXiv:1711.10566.

Raissi, M., Perdikaris, P. & Karniadakis, G. E. (2017b) Physics informed deep learning (part I): data-driven solutions of nonlinear partial differential equations. arXiv:1711.10561.

Raissi, M., Yazdani, A., Karniadakis, G. E. (2018) , Hidden fluid mechanics: A Navier-Stokes informed deep learning framework for assimilating flow visualization data - arXiv preprint arXiv:1808.04327, 2018

Raissi, M., Perdikaris, P. & Karniadakis, G. E. (2019a)  Physics-informed neural networks: A deep learning framework for solving forward and inverse problems involving nonlinear partial differential equations, Journal of computational Physics, Volume 378, 2019, Pages 686-707, ISSN 0021-9991. https://doi.org/10.1016/j.jcp.2018.10.045.

Raissi, M., Wang, Z., Triantafyllou, M., & Karniadakis, G. E. (2019b). Deep learning of vortex-induced vibrations. Journal of Fluid Mechanics, 861, 119-137. doi:10.1017/jfm.2018.872

Singh, A. P., and Duraisamy, K.,  "Using field inversion to quantify functional errors in turbulence closures," Physics of Fluids, Vol. 28, No. 4, pp. 045110, (2016). https://aip.scitation.org/doi/abs/10.1063/1.4947045

Sekar, V., Jiang, Q., Shu, C., and Khoo, B.C., Fast flow field prediction over airfoils using deep learning approach Physics of Fluids 31, 057103 (2019).

Thuerey, N., Weißenow, K., Prantl, L. and Hu, X. Deep Learning Methods for Reynolds-Averaged Navier–Stokes Simulations of Airfoil Flows, AIAA Journal 2020 58:1, 25-36

Weller, H. G., Tabor, G., Jasak, H., and Fureby, C., "A tensorial approach to computational continuum mechanics using object-oriented techniques," Computers in Physics, Vol. 12, No. 6, Nov/Dec, (1998).

Wang, Z., Luo, K., Li, D., Tan, J., and Fan, J., "Investigations of data-driven closure for subgrid-scale stress in large-eddy simulation," Physics of Fluids, Vol. 30, No. 12, pp. 125101, (2018). https://aip.scitation.org/doi/abs/10.1063/1.5054835

White, F.M., "Viscous Fluid Flow", 3rd Edition, McGraw-Hill, Boston, (2006).

Wu, J-L., Xiao, H., and Paterson, E., "Physics-informed machine learning approach for augmenting turbulence models: A comprehensive framework," Phys. Rev. Fluids 3, 074602, (2018).

Zhu, L., Zhang, W., and Kou, J., and Liu,Y., "Machine learning methods for turbulence modeling in subsonic flows around airfoils," Physics of Fluids, Vol. 31, No.1, (2019).